\documentclass[twocolumn]{article}
\usepackage[utf8]{inputenc}

\usepackage{geometry}
\usepackage{soulutf8}

\usepackage{caption}
\usepackage{subcaption}
\usepackage{booktabs}
\usepackage{adjustbox}
\usepackage{multirow}
\usepackage[table]{xcolor}
\usepackage{floatrow}

\usepackage{amsmath}
\usepackage{amssymb}
\usepackage{bm}
\usepackage{nicefrac}
\usepackage{mathtools}

\usepackage[separate-uncertainty=true,parse-numbers=true]{siunitx}

\usepackage{pgfplots}
\usepackage{tikz}
\pgfplotsset{compat = newest}
\usetikzlibrary{quotes,angles,calc,arrows,patterns,quotes,external}
\usepgfplotslibrary{groupplots}
\usepackage[hidelinks]{hyperref}
\usepackage[capitalise,nameinlink,noabbrev]{cleveref} % after amsmath
\usepackage[nocompress]{cite}

\usepackage{authblk}

\renewcommand{\vec}{\boldsymbol}
\newcommand{\mat}{\mathbf}
\newcommand{\trans}{^{\mathrm{T}}}

\definecolor{crimson2143940}{RGB}{214,39,40}
\definecolor{darkgray176}{RGB}{176,176,176}
\definecolor{darkorange25512714}{RGB}{255,127,14}
\definecolor{forestgreen4416044}{RGB}{44,160,44}
\definecolor{mediumpurple148103189}{RGB}{148,103,189}
\definecolor{sienna1408675}{RGB}{140,86,75}
\definecolor{steelblue31119180}{RGB}{31,119,180}

\title{Impact Study of Numerical Discretization Accuracy on Parameter
  Reconstructions and Model Parameter Distributions}

\author[a]{Matthias Plock}
\author[a,b]{Martin Hammerschmidt}
\author[a,b,$\dagger$]{Sven Burger}
\author[a,b]{Philipp-Immanuel Schneider}
\author[a]{Christof Sch\"utte}
\affil[a]{Zuse Institute Berlin, Takustraße 7, 14195 Berlin, Germany}
\affil[b]{JCMwave GmbH, Bolivarallee 22, 14050 Berlin, Germany}
\affil[$\dagger$]{Corresponding author: burger@zib.de}
\date{June 21, 2023}

\begin{document}

\maketitle

\begin{abstract}
  In optical nano metrology numerical models are used widely for parameter
  reconstructions. Using the Bayesian target vector optimization method we fit a
  finite element numerical model to a Grazing Incidence X-Ray fluorescence data
  set in order to obtain the geometrical parameters of a nano structured line
  grating. Gaussian process, stochastic machine learning surrogate models, were
  trained during the reconstruction and afterwards sampled with a Markov chain
  Monte Carlo sampler to determine the distribution of the reconstructed model
  parameters. The numerical discretization parameters of the used finite element
  model impact the numerical discretization error of the forward model. We
  investigated the impact of the polynomial order of the finite element ansatz
  functions on the reconstructed parameters as well as on the model parameter
  distributions. We showed that such a convergence study allows to determine
  numerical parameters which allows for efficient and accurate reconstruction
  results.
\end{abstract}

\noindent{\it Keywords\/}: Bayesian target vector optimization, Parameter
reconstruction, Model accuracy, Least squares, Bayesian optimization, Markov
chain Monte Carlo.

\section{Introduction}
\label{sec:introduction}

In the field of optical nano metrology dimensions of samples which are
structured at the nanoscale are quantified by observing how the sample interacts
with impinging light of precisely defined
properties~\cite{diebold2013nanoscale}. This is of particular interest in
semiconductor manufacturing, where feature sizes need to be controlled on a
level much smaller than the wavelength of light in the visible
spectrum~\cite{orji2018metrology, denboef2016optical, endres2014investigations}.
Various setups are used, where, e.g., the interaction of light with the sample
is measured and analyzed at different wavelengths, angles of incidence,
polarizations, etc.
\cite{jones2003small,omullane2016modeling,attota2016feasibility,Soltwisch2016prb}.
As the measured spectra do not directly reveal the geometrical parameters, a
common solution is to create a parameterized model of the measurement process.
This model is also referred to as a forward problem. Assuming that the model
accurately represents the measurement process, it can then be fit to the
experimental measurement. The parameters resulting from the fit can then be used
to indirectly explain the, e.g., geometrical parameters of the investigated
structure~\cite{aster2018parameter}. Many different fitting methods have been
shown to be capable of solving this inverse (or parameter reconstruction)
problem. These include local methods such as the Gauss-Newton scheme or the
Levenberg-Marquardt algorithm~\cite{Hamm:17,STORCH20071417}, Nelder-Mead or the
limited memory Broyden–Fletcher–Goldfarb–Shanno algorithm with box constraints
(\mbox{L-BFGS-B})~\cite{Schn:2019Benchmark}, or heuristic global methods like
particle swarm optimization~\cite{PSO_magnetotelluric,schwaab2008nonlinear} or
differential evolution~\cite{lobato2012estimation,cavalini2016model}. One can
also maximize the appropriate likelihood function using Markov chain Monte Carlo
(MCMC) sampling methods~\cite{Herrero:21}.

Due to the generally complex nature of the measurement process, the
parameterized models are typically implemented using numerical methods. Such
numerical models have two sets of parameters: a set of \textit{physical}
parameters which encode physical quantities (such as geometrical parameters),
and a set of \textit{numerical} parameters which control the numerical
approximation accuracy. Selecting appropriate numerical parameters is an
integral part of the creation of numerical models. They are usually determined
by means of a convergence study. Here, the numerical parameters are
systematically varied while the physical parameters are kept fixed. The model
output is then compared to a reference solution which is assumed to be very
accurate. The numerical parameters are chosen such that (i) a convergence trend
of the output can be seen and (ii) a desired accuracy is reached. Appropriate
numerical parameters can trade off numerical accuracy and computational costs,
but more importantly assert that the results of the model are reliable.

\begin{figure}[ht]
  \centering
  \includegraphics{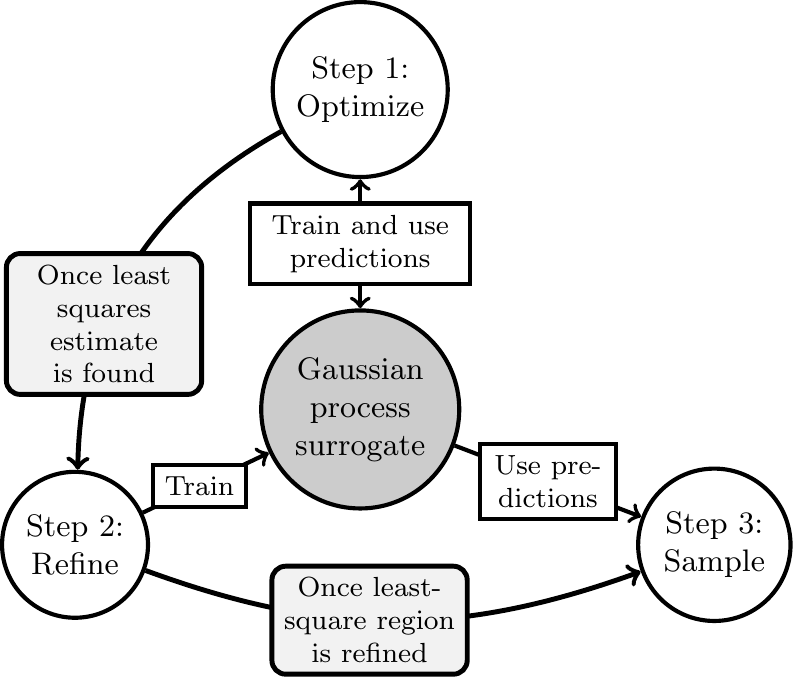}
  \caption{A schematic of the complete three-step reconstruction process. At
    each step a Gaussian process surrogate model is trained and/or its
    predictions are used. In Step 1, the Bayesian target vector optimization
    scheme is applied and a least squares estimate of the forward problem is
    determined. In Step 2, the surrogate model trained during Step 1 is refined,
    and the fidelity in the region around the found least squares estimate is
    increased. Finally, the full model parameter distribution is determined in
    Step 3 by sampling the refined surrogate model with a Markov chain Monte
    Carlo method. }
  \label{fig:schematic}
\end{figure}

In this article we extend this approach. Additionally, we first determine a
least squares estimate by performing complete parameter reconstructions, using a
set of different numerical parameters. For a subset of the considered numerical
parameters we then determine the full (potentially non-Gaussian) model parameter
distribution by means of a MCMC sampling method. We finally investigate the
dependence of the reconstructed parameters on the numerical parameters.
Na\"ively this is prohibitively expensive for any model for which the evaluation
requires more than a few seconds of wall time. The task becomes feasible by
using an approach centered around Gaussian processes
(GPs)~\cite{williams2006gaussian}. GPs are machine learning \emph{surrogate}
models that can be trained on observations of the actual model. After they are
trained they can serve as a cheap-to-evaluate predictor for the actual model,
and can therefore also be used as an approximation of them. We use the Bayesian
target vector optimization scheme~\cite{plock2022bayesian} for performing the
parameter reconstructions. Afterwards we apply a MCMC sampler to infer the full
model parameter distribution from the GP surrogates trained during the parameter
reconstruction. For an accurate determination of the full model parameter
distribution, the surrogate model has to be a faithful representation of the
actual model close to the found least squares estimate. Therefore, prior to the
MCMC sampling, a surrogate refinement stage is entered in which the fidelity of
the GP surrogates is increased. This is achieved by adding observations of the
actual model in the region to be sampled by the MCMC sampler. The complete
parameter reconstruction workflow including MCMC sampling is sketched in
\cref{fig:schematic}.

This article is organized as follows. \cref{sec:theory} contains the theoretical
background. We discuss parameter reconstructions and how to find a least squares
estimate, which includes a simple estimate for model parameter uncertainties
based on the Jacobian matrix. We then review the BTVO method and discuss how the
GP surrogates trained during a BTVO run can be used for MCMC sampling. Finally,
we briefly discuss ways of controlling the accuracy of the numerical model. In
\cref{sec:experiments} we investigate the impact of a proper or improper choice
of numerical parameters. We revisit a parameter reconstruction problem, in which
a line grating is investigated in a Grazing Incidence X-Ray Fluorescence (GIXRF)
experiment~\cite{andrle2021shape, plock2022bayesian}. We first perform a
conventional convergence analysis. For each of the numerical parameters
considered in the convergence analysis we then determine a least squares
estimate by performing a parameter reconstruction using the BTVO. Finally, for a
subset of the numerical parameters considered we determine the full model
parameter distribution using the GP surrogate aided MCMC approach.

\section{Background}
\label{sec:theory}

We briefly review the theoretical fundamentals for parameter reconstructions
using the Bayesian target vector optimization method~\cite{plock2022bayesian}.
Moreover, we briefly discuss ways of controlling the level of accuracy in finite
element method (FEM) simulations.

\subsection{Parameter reconstructions and the least squares problem}
\label{subsec:least_square}

Parameter reconstruction problems in optical metrology are often least squares
type problems. The task here is to fit the $K$ ouputs of a parameterized forward
model function $\vec{f}(\vec{p})$ to a set of experimental measurements $\vec{t}
= \left( t_{1}, \dots, t_{K} \right)\trans$, with measurement uncertainties
$\vec{\eta} = \left( \eta_{1}, \dots, \eta_{K} \right)\trans$, model parameters
$\vec{p} \in \mathcal{X} \subset \mathbb{R}^{N}$, and forward model $\vec{f}:
\mathcal{X} \to \mathbb{R}^{K}$. We assume that the forward model function is a
good approximation of the measurement process, and that model errors can
therefore be neglected. Furthermore the forward model function is treated as a
black box that simply maps input parameters $\vec{p}$ to output values $\vec{y}
= \vec{f}(\vec{p})$. Here we assume that $\vec{f}$ is once differentiable and
that each of the $K$ outputs can be modeled by a Gaussian process (GP).

For a \emph{true} parameter $\vec{p}_{t}$ we assume that the output of the model
$\vec{f}(\vec{p}_{t})$ equals the experimental dataset plus a Gaussian noise
contribution with zero-mean and variance $\eta_{k}^{2}$, coming from the
measurement uncertainty, i.e.,
\begin{equation}
  \label{eq:calc_model_output}
  t_{k} = f_{k}(\vec{p}_t) + \varepsilon_{k} \quad \text{with} \quad
  \varepsilon_{k} \sim \mathcal{N}(0, \eta_{k}^2) \,.
\end{equation}
When the measurement uncertainty itself is unknown one can employ fully Bayesian
methods that determine maximum likelihood or posterior estimates of the
parameters\cite{martino2021automatic}. Finding a good least squares estimate
(LSQE) for the unknown true parameter can be considered an optimization problem.
This is often solved by minimizing the residual sum of squares
\begin{equation}
  \label{eq:sum_sq_res}
  \chi^2(\vec{p}) = \sum_{k=1}^{K} \frac{\left( f_{k}(\vec{p}) - t_{k}
    \right)^2}{\eta_{k}^2} \,,
\end{equation}
i.e., by finding
\begin{equation}
  \label{eq:p_lsqe}
  \vec{p}_{\mathrm{LSQE}} = \underset{\vec{p} \in \mathcal{X}}{\arg\min}\,
  \chi^2(\vec{p}) \,.
\end{equation}
A solution for \cref{eq:p_lsqe} can be found, e.g., by using least squares
methods such as Levenberg-Marquardt~\cite{leve:1944,marq:1963,flet:1971} or the
Bayesian target vector optimization (BTVO) method presented in
\cref{subsec:bayes_lsq}.

Further, also the uncertainties of the model parameters are of interest. For
least squares methods such as Levenberg-Marquardt one can use the Jacobian
matrix at the least squares estimate to calculate the covariance matrix
$\mat{Cov}(\vec{p}_{\mathrm{LSQE}})$, which can then be used to determine
Gaussian uncertainty bands $\vec{\epsilon}_{\mathrm{LSQE}}$, i.e., uncertainties
in terms of symmetric $1\sigma$ intervals around
$\vec{p}_{\mathrm{LSQE}}$~\cite{kutner2005applied,strutz2010data,press2007numerical}.

\subsection{A Bayesian optimization approach to parameter reconstructions}
\label{subsec:bayes_lsq}

Bayesian optimization (BO) methods~\cite{brochu2010tutorial,jones1998efficient}
are sequential optimization methods that can be very efficient for globally
optimizing expensive black box functions
\cite{jones1998efficient,Schn:2019GPR,Schn:2019Benchmark}. At each iteration
during the optimization process a stochastic machine learning surrogate model --
most often a GP~\cite{williams2006gaussian,garnett_bayesoptbook_2023} -- is
trained using all previously observed values of the forward model. GPs are
usually much cheaper to evaluate than the model function itself.

By considering all previous forward model observations seen during the
  optimization a global model of the expensive forward model function is
created. This gives BO methods an advantage over conventional optimization
methods, that often only create a local model of the forward model, e.g., in the
form of a Hessian (e.g., for L-BFGS-B~\cite{byrd1995limited}) or Jacobian (e.g.,
Gauss-Newton type methods~\cite{deuflhard2005newton}) at the point of
evaluation.

In BO the optimization typically happens in a cyclic fashion that repeats in
each step of the optimization. First the GP surrogate is trained on the
observations of the model function that have been seen in all previous
iterations. Afterwards the length scales of the GP are optimized such that they
best describe the training data. This \emph{hyperparameter optimization} is
usually computationally rather expensive and can also be done sparsely, e.g.,
every $M_{\mathrm{Hyp}}$ iterations. Now the GP surrogate can be used as a
predictor or approximator of the training data. Finally, the next sampling
position $\vec{p}_{m+1}$ is selected by maximizing an acquisition (or utility)
function. This acquisition function uses the predictions made by the GP in order
to achieve the goal of the optimization. This new sample $\vec{p}_{m+1}$ and the
associated black box function value $f(\vec{p}_{m+1})$ are then added to the
training data set for the next iteration.

Each iteration in the optimziation  contains multiple smaller
optimization steps. This is numerically expensive and the reason why BO methods
are usually reserved for model functions for which the calculation of point
values is at least as expensive as one of the nested optimization cycles.

This approach can be adopted to solve least squares parameter reconstruction
problems. In the References~\cite{plock2022bayesian,uhrenholt} $K$ independent
GPs are trained on the $K$ outputs of the vector valued forward model function
$\vec{f}(\vec{p})$. At each iteration in the optimization an appropriate utility
function then generates new sample candidates that can lead to an improvement
over the currently best $\chi^2$ value. The Bayesian target vector optimization
(BTVO) scheme~\cite{plock2022bayesian} implements several key performance
improvements that make the method perform well even for a large number $K$ of
components in the forward model function $\vec{f}(\vec{p})$.

In the following we will first briefly review GPs as essential building blocks
for the BTVO scheme, and then give an overview over the scheme itself and the
improvements over the approach taken by Uhrenholt and Jensen~\cite{uhrenholt}.

\subsubsection{Gaussian processes}
\label{subsec:gps}

A Gaussian process (GP)~\cite{williams2006gaussian,garnett_bayesoptbook_2023} is
a stochastic machine learning surrogate model that is defined on the continuous
domain $\mathcal{X}$. GPs can best be understood as an extension of finite
dimensional multivariate normal distributions to an infinite dimensional case. A
multivariate normal distribution is completely specified by a mean vector
$\vec{\mu}$ and a covariance matrix $\mat{\Sigma}$. For a GP these are replaced
by a continuous mean function $\mu: \mathcal{X} \to \mathbb{R}$ and a covariance
kernel function $k: \mathcal{X} \times \mathcal{X} \to \mathbb{R}$. Frequent
choices~\cite{Schn:2019Benchmark} which are also applied in this article are a
constant mean function $\mu(\vec{p})~=~\mu_{0}$ and a Mat\'ern $\nicefrac{5}{2}$
covariance kernel function~\cite{brochu2010tutorial}, i.e.,
\begin{gather}
  k(\vec{p}, \vec{p}^{\prime}) = \sigma_{0} \left( 1 + \sqrt{5}r +
    \frac{5}{3}r^2 \right) \times \exp{\left( -\sqrt{5}r \right)} \,,\\
  \text{where} \quad r = \sqrt{\sum_{n=1}^{N} \left( \frac{ p_{n} -
        p^{\prime}_{n} }{l_{n}} \right)^{2}} \,.
\end{gather}
The quantity $r$ gives the normalized distance between the two model parameter
vectors $\vec{p}$ and $\vec{p}^{\prime}$. GPs are parameterized by a set of
hyperparameters $\{\mu_{0}, \sigma_{0}, l_{1}, \dots, l_{N}\}$ (each one $\in
\mathbb{R}$), where $\sigma_0$ is the prior standard deviation and $l_{n}$ are
the length scales of the forward model parameters of the GP. These are chosen to
maximize the likelihood of the observations. Once a GP has been trained on $M$
observations of a function $f: \mathcal{X} \to \mathcal{Y} = \mathbb{R}$, i.e.,
on $\vec{Y} = \left[ f(\vec{p}_{1}), \dots, f(\vec{p}_{M}) \right]\trans$, it
can be used to predict a normal distribution at each point $\vec{p}_{\ast}$ in
the parameter space,
\begin{gather*}
  \hat{f}(\vec{p}_{\ast}) \sim \mathcal{N}(\overline{y}(\vec{p}_{\ast}),
  \sigma^{2}(\vec{p}_{\ast})) \,.
\end{gather*}
The mean of the normal distribution approximates the training data, while the
predicted variance encodes, how much the training data can be expected to
scatter around the predicted mean -- given the chosen hyperparameters. Here we
have used the hat-notation (i.e. $\hat{f}(\vec{p}_{\ast})$) to denote a random
variable. For the parameter dependent predicted mean
$\overline{y}(\vec{p}_{\ast})$ and variance $\sigma^{2}(\vec{p}_{\ast})$ of the
GP we have
\begin{align*}
  % \label{eq:pred_mean}
  \overline{y}(\vec{p}_\ast) &= \mu_0 + \vec{k}\trans (\vec{p}_\ast)
                               \mat{K}^{-1}[\vec{Y}-\mu_0 \vec{1}] \,,\\
                               % \label{eq:pred_var}
  \sigma^2(\vec{p}_\ast) &= \sigma_0^2 - \vec{k}\trans (\vec{p}_\ast)
                           \mat{K}^{-1} \vec{k}(\vec{p}_\ast) \,,
\end{align*}
where the covariance kernel function $k(\vec{p}, \vec{p}^{\prime})$ is used to
compute $\vec{k}(\vec{p}_\ast) = \left[ k(\vec{p}_\ast,\vec{p}_1), \dots,
  k(\vec{p}_\ast,\vec{p}_{M}) \right] \trans$, $(\mat{K})_{i j} =
k(\vec{p}_i,\vec{p}_j)$, and $\mat{1}$ denotes a $M\times M$ identity matrix.

\subsubsection{Bayesian target vector optimization}
\label{subsec:btvo}

In order to solve the problem of determining $\vec{p}_{\mathrm{LSQE}}$ the BTVO
scheme trains $K$ independent GP surrogates with the $K$ outputs of the
parameterized forward model $\vec{f}(\vec{p})$. An alternative approach is to
train a single multi-output GP~\cite{alvarez2011kernels,liu2018remarks} that is
capable of predicting all $K$ outputs at the same time. Na\"ively the size of
the covariance matrix grows by a factor of $K$ at each iteration. For models
with a small number of channels this can be a viable option. For forward models
with $K \sim 100$ or larger this can however present an issue, as the covariance
matrix has to be inverted at each iteration at a cost of approximately
$\mathcal{O}(K^3)$ FLOPS~\cite{press2007numerical}. This high complexity can be
reduced drastically by e.g. choosing a separable covariance
kernel~\cite{alvarez2011kernels}. It has successfully been applied in
Reference~\cite{matsui2019bayesian}, where a separable kernel was employed to
account for correlations between model output channels in a least squares
regression problem.

Na\"ively, the hyperparameters of each GP have to be optimized individually. In
the BTVO scheme this is avoided by sharing the covariance matrix between
all $K$ GPs. This approach can be motivated by the assumption that in many
of the problems considered one performs, e.g., an angular or
wavelength/energy scan for the same model parameters. Here one can assume that
the length scales of each of the scanned parameters are similar.
Accordingly, at each iteration, only one comparatively small
covariance matrix has to be inverted instead of $K$. The predicted normal
distributions of each GP are then combined in the calculation of a stochastic
chi-square predictor variable $\hat{\chi}^2(\vec{p})$, i.e.,
\begin{equation}
  \label{eq:pred_chisq}
  \hat{\chi}^2(\vec{p}) = \sum_{k=1}^{K}\frac{\left( \hat{f}_{k}(\vec{p}) - t_{k} \right)^2}{\eta_{k}^2} \,.
\end{equation}
Since this random variable is composed of a sum of non-central normal
distributions with non-unit variance, it follows a \emph{generalized} chi-square
distribution. In principle it is possible to calculate values for this
distribution, this can however be very
costly~\cite{mathai1992quadratic,mohsenipour2012distribution}. In order to
efficiently incorporate this distribution into the optimization scheme it is
therefore approximated twice, such that its values are ultimately given by a
parameterized normal distribution. This allows for an efficient implementation
of the lower confidence bound acquisition function~\cite{uhrenholt} employed in
the BTVO method.

Unfortunately, these approximations introduce numerical issues within the
acquisition function. For a large number of channels $K$ in the forward model
the probability of finding chi-square values smaller than the current
optimum very rapidly tends towards zero as the number of iterations
increases~\cite{plock2022bayesian}. This reduces the exploratory nature of
the scheme quite drastically, as the acquisition function will generally only
generate sample candidates close to previously sampled positions. In order to
restore the exploratory nature of the scheme the BTVO  uses a
modified parameterization of the approximated probability
distribution~\cite{plock2022bayesian}.

The BTVO method has been shown to be capable of outperforming established
parameter reconstruction methods, and that it can reach reconstructed parameters
in fewer iterations than, e.g.,
Levenberg-Marquardt~\cite{plock2022bayesian}. The method has since
been applied to other expensive parameter reconstruction
problems~\cite{schneider2022reconstructing}.

Naturally, one could use the conventional BO method to solve the
least squares problem directly~\cite{Schn:2019Benchmark}.
The BTVO scheme has two major advantages over this na\"ive approach. First, by
training a stochastic model for each component of the model function a lot of
information is retained about the interplay of the $K$ channels of the model
function. By determining the residual \emph{sum} of \emph{squares} (where both
summing up and squaring erase information) invaluable information is lost that
can provide one with an advantage during optimization. And second, the employed
chi-square distribution used in the BTVO better matches the minimized chi-square
value. During a parameter reconstruction with the conventional BO a single GP is
trained on chi-square values (defined on $\mathbb{R}^{+}$, including zero) that
predicts a normal distribution (defined on $\mathbb{R}$). In regions where
this chi-square training data approaches zero it can happen that
the predictions of the GP will be wrong or at least inaccurate. Critically, this
is precisely the case when the goal of the optimization of finding parameter
values with small chi-square values is reached. Both of these issues lead to
sub-optimal optimziation performance.

\subsection{Efficient determination of the model parameter distribution via
  Markov chain Monte Carlo}
\label{subsec:parameter_uncertainties}

Since derivatives of the forward model $\vec{f}(\vec{p})$ w.r.t. all model
parameters are easily calculated using GPs~\cite{garcia2018shape} it is possible
to, similar to, e.g., Levenberg-Marquardt, determine Gaussian
uncertainty bands $\vec{\epsilon}_{\mathrm{LSQE}}$ by means of the covariance
matrix $\mat{Cov}(\vec{p}_{\mathrm{LSQE}})$. Linear correlations between the
model parameters are also easily
revealed~\cite{kutner2005applied,strutz2010data}.

A much more informative quantity is the \emph{full} model parameter
distribution, from which one can extract the model parameter uncertainties in
terms of \SI{16}{\percent}, \SI{50}{\percent} (i.e., the median), and
\SI{84}{\percent} percentiles, as well as potentially non-linear correlations
between the various model parameters. The full model parameter distribution can
generally not be established in an analytical fashion. Instead it is often
determined by means of sampling methods, such as a Markov chain Monte Carlo
(MCMC)~\cite{andrieu2003introduction,sammut2011encyclopedia,friedman2001elements}
ansatz.

For parameter reconstruction problems, the MCMC sampler draws samples from the
likelihood function for $K$ independent normally distributed random
variables~\cite{degroot2012probability}
\begin{equation}
  \label{eq:least_square_likelihood}
  \mathcal{L}(\vec{p}) = \prod_{k = 1}^{K} \frac{1}{\sqrt{2\pi\eta_{k}^2}}
  \exp{ \left( - \frac{1}{2} \frac{\left( f_{k}(\vec{p}) - t_{k} \right) ^2}{\eta_{k}^{2}} \right) } \,,
\end{equation}
and uses them to construct one or more Markov chains, i.e., chains of samples
from the domain of the sampled forward model $\vec{f}(\vec{p})$. The
equilibrium distribution of the samples in these chains resemble the desired
model parameter distribution. This distribution can, e.g., be visualized and
analyzed by constructing histograms of the the samples in the chains.

MCMC allows to even extend the least squares ansatz of
constant measurement errors~\cite{strutz2010data}, and instead fit a more
complex error model to the measured data. It is therefore an invaluable method
in many areas of science and engineering, not just for metrology applications.

In order to construct the equilibrium distribution the likelihood function is
evaluated often several tens of thousands of times. This can make the
method unattractive from a resource standpoint if the forward model
$\vec{f}(\vec{p})$ in the likelihood function is expensive to evaluate.
Using a trained surrogate model instead of the actual forward model can lessen
the resource impact drastically, since their evaluation is often orders of
magnitude faster than the forward model.
Reference~\cite{plock2022bayesian} proposes to use the GP
surrogates trained during an optimization run with the BTVO scheme.

An issue that arises when training these surrogate models is that one has to
assert that the model possesses a sufficient degree of fidelity in the
investigated region, i.e., around the least squares
estimate $\vec{p}_{\mathrm{LSQE}}$. For the surrogates trained during a BTVO run
we can assume that they are trained in the region from which the
least squares estimate was found. To ensure that the
entire region is trained sufficiently well, a refinement stage is entered in
which parameter samples are drawn from the normal distribution
$\mathcal{N}(\vec{p}_{\mathrm{LSQE}}, \mat{Cov}(\vec{p}_{\mathrm{LSQE}}))$ and
used to evaluate the forward model $\vec{f}(\vec{p})$. The results are used to
train the surrogate further. This refinement process is stopped once the
predicted variance of the GPs is below a certain
threshold~\cite{plock2022bayesian}.

The actual MCMC sampling of the GP surrogate of the forward model is performed
post-refinement using \texttt{emcee}~\cite{emcee}. The created chains are
analyzed and visualized using \texttt{corner.py}~\cite{cornerpy}, thereby
revealing the individual marginal model parameter distributions, its
\SI{16}{\percent}, \SI{50}{\percent} (i.e., the median), and \SI{84}{\percent}
percentiles, as well as potential non-linear correlations between the model
parameters.

\subsection{Numerical solution of the forward model}
\label{subsec:fem}

The parameterized forward model $\vec{f}(\vec{p})$ maps input points $\vec{p}$
from the parameter space to output values $\vec{y}$. These are obtained by
solving the linear, time-harmonic Maxwell's equations for scattering problems.
In general, solutions for this type of problem cannot be found analytically.
Therefore a numerical solver based on the finite element method
(FEM)~\cite{Monk2003a} is employed for obtaining solutions. The solver used in
this article is \texttt{JCMsuite}~\cite{jcm_scattering, Burger2008ipnra}.

The geometry of the investigated scattering problem is discretized in space and
thus approximated by a mesh of non-overlapping discrete elements (e.g.,
triangles for a problem in two dimensions). On these discrete elements the
fields of Maxwell's equations are expanded into a set of polynomial ansatz
functions. This allows to formulate the problem of finding a solution to the
scattering problem in terms of algebraic equations, which can be tackled
efficiently using numerical methods.

Two ways of controlling the numerical accuracy of the solution of the problem
are (i) controlling the size $h$ of the discrete elements used in the mesh, and
(ii) varying the number or degree $p$ of the polynomial ansatz functions. Both,
reducing the size of the elements $h$ and increasing the polynomial degree $p$
increases the resolution of the field expansion, i.e., reduces the numerical
discretization error. This naturally also increases the resource demand, since
more quantities need to be calculated when finding a solution. It is therefore
important to find a balance between the numerical parameters $h$ and $p$, and
the required accuracy of the solution.

\section{Computer experiments}
\label{sec:experiments}

We have used the Bayesian target vector optimization (BTVO) method
and the associated surrogate aided Markov chain Monte Carlo (MCMC)
sampling approach to investigate the influence of the accuracy of an underlying
finite element method (FEM) model on a parameter reconstruction.
Here, the accuracy of the FEM model was controlled by varying the polynomial
degree $p$.

The parameter reconstruction is performed by fitting the FEM model to an
experimental data set. This data set was recorded during a Grazing Incidence
X-Ray Fluorescence (GIXRF) experiment, performed by the PTB at the BESSY II
light source in Berlin~\cite{C8NR00328A}. The measurement process and the FEM
model is briefly discussed in \cref{subsec:model}. We have previously reported
on the technical details on the BTVO method and used the same data set and model
to demonstrate the method~\cite{plock2022bayesian}.

In our computer experiments we conducted a convergence study of the FEM model,
where we observed how the simulated output of the model changes for different
polynomial degrees $p \in \{2, \dots, 7\}$. We compared each result to a
reference solution obtained using $p_{\mathrm{ref}} = 8$.

For each of the polynomial degrees $p$ considered in the convergence study we
then performed a parameter reconstruction benchmark. Here, six independent
parameter reconstructions using the BTVO method were performed for each $p$. The
results at each iteration were compared to a trusted reference solution.

Finally, we have refined a small subset of the surrogates for different $p$
created during the optimization benchmarks. Using these refined surrogates we
have then determined the full model parameter distribution using the surrogate
aided MCMC sampling approach.

All experiments were performed on the High Throughput Computing cluster at the
Zuse Institute Berlin, on a Dell PowerEdge C6520 rack server with two Intel Xeon
Gold 6338 processors, yielding 64 cores or 128 threads. The nodes were equipped
with 1024 GB of RAM. The simulations were set up to use all available threads.
The \texttt{JCMsuite} version employed was 5.2.4. The BTVO
  scheme used in the experiments was part of the optimizer that was
shipped with the solver.

\subsection{The experimental dataset and the FEM model}
\label{subsec:model}

Grazing Incidence X-Ray Fluorescence (GIXRF) \cite{andrle2021shape} is a
destruction free indirect optical measurement method that can be used to
determine geometrical as well as material parameters of periodically
nano-structured samples. The investigated grating is illuminated with
monochromatic light in the X-ray regime, which penetrates the sample material to
some extent. The reflected light interferes with the incident light and forms an
X-ray standing wave field, which can locally increase light-matter interaction
within the sample material. A portion of the incident energy is absorbed by the
sample, but most is ultimately given off again in the form of a fluorescence
spectrum. This is integrated and recorded by a calibrated silicon drift
detector. The recorded signal depends very strongly on the incident angle, the
sample material, as well as the sample geometry.

Exploiting the strong angular dependence of the fluorescence signal, an angular
scan is performed over \num{208} discrete incidence angles, using X-ray light
with an energy of \SI{520}{\electronvolt}.

To reconstruct the experimental conditions and geometrical parameters of the
periodic sample, a parameterized FEM model of the measurement process is created
using the Maxwell solver \texttt{JCMsuite}. In it the electromagnetic fields of
the X-ray standing wave field are simulated on a periodic unit cell of the
geometry, and the modified 2D Sherman equation is subsequently used to calculate
the fluorescence signal. This is done for each of the angles used in the angular
scan of the experiment. The model function contains ten free parameters. Of
these, seven are geometrical (e.g., the critical dimension), two are there to
account for uncertainties in the incidence angles, and one is used to scale the
simulated intensities to the experimental intensities.

We have previously reported on the data set~\cite{plock2022bayesian}. For more
information about the numerical model, a plot of an an exemplary unit cell
geometry, the complete parameters and parameter ranges, and the measured GIXRF
signal, we therefore refer the reader to
Reference~\cite{plock2022bayesian}.

\subsection{Model convergence}
\label{subsec:model_convergence}

\begin{figure*}[ht]
  \centering
  \begin{subfigure}[b]{0.59\textwidth}
    \centering
    \includegraphics{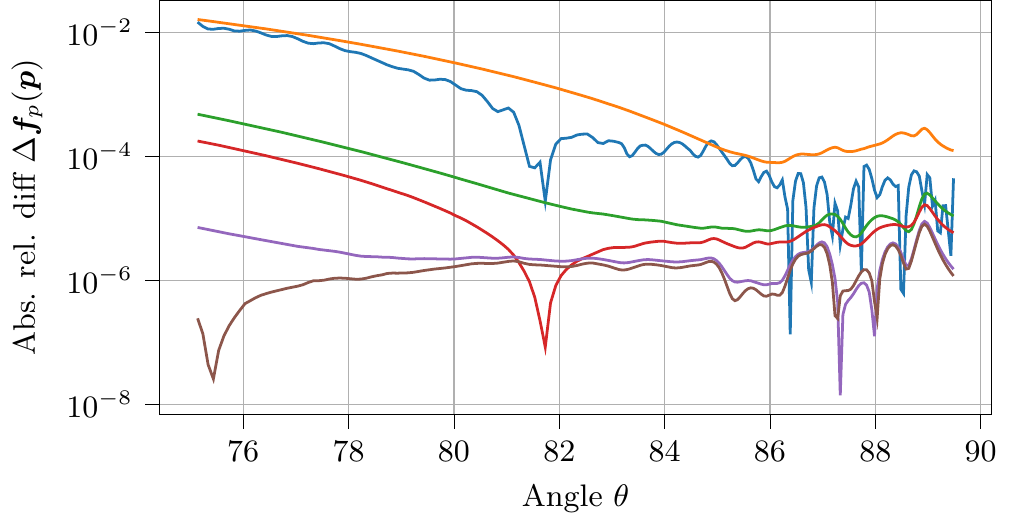}
  \end{subfigure}
  \hfill
  \begin{subfigure}[b]{0.39\textwidth}
    \centering
    \includegraphics{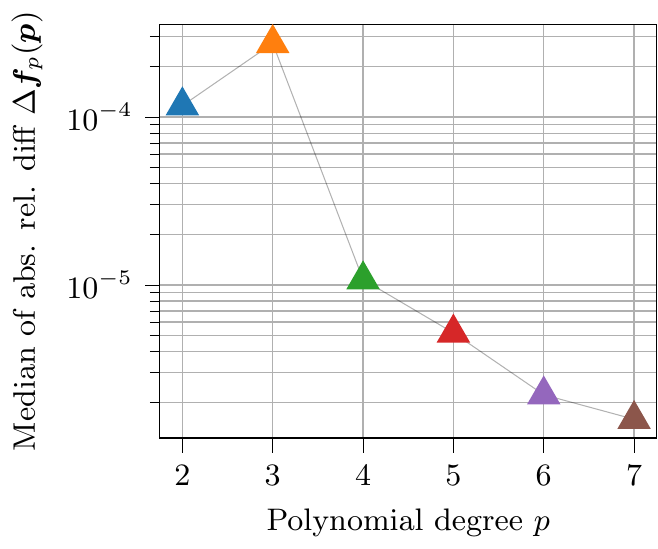}
  \end{subfigure}
  \caption{Convergence plot of the model for a random parameter $\vec{p}$ in the
    parameter space. Shown is the absolute relative difference between the
    calculated model output for a polynomial degree $p \in \{ 2, \dots, 7\}$ and
    a reference model output with $p_{\mathrm{ref.}} = 8$. The
    model outputs for $p = 2$ and $p = 3$ can not be considered converged. For
    $p = 2$ the model output fluctuates wildly w.r.t. the reference solution,
    and for $p = 3$ the difference from the reference solution is very large
    compared to the remainder of the considered polynomial orders. The absolute
    relative difference is defined as $\Delta \vec{f}_{p}(\vec{p}) =
    \nicefrac{\left| \vec{f}_{p}(\vec{p}) - \vec{f}_{\mathrm{ref}}(\vec{p})
      \right| }{\vec{f}_{\mathrm{ref}}(\vec{p})}$, where division by the
    reference vector is meant component-wise. The right sub-figure serves as a
    color guide for the figures in this article.}
  \label{fig:convergence}
\end{figure*}

In the model convergence analysis we investigate the convergence of the
simulated fluorescence intensities for difference polynomial degrees $p$, for
each of the angles $\theta$ considered in the experiment. To do this, a random
model parameter $\vec{p}$ from the allowed parameter domain was generated and
used to evaluate the different $p$ discretizations of the FEM model with. We
chose $p \in \{ 2, \dots, 7\}$, and compared the simulated fluorescence
intensities to a reference solution generated with a FEM model with
$p_{\mathrm{ref.}} = 8$. The largest allowed side length of an element in the
geometry discretization was set to $h = \SI{5}{\nano\meter}$. At this value of
$h$ we have observed that the rounded parts of the mesh, i.e. parts
parameterized by the parameter $R$, were resolved reasonably well. Varying $p$
instead of $h$ allowed us to efficiently vary the accuracy of the FEM model
without compromising the quality of the geometry representation.

The convergence is determined by calculating the absolute value of the relative
difference to the reference solution, i.e., as
\begin{equation}
  \label{eq:absreldiff}
  \Delta f_{p}(\vec{p}) = \frac{\left| f_{p}(\vec{p}) - f_{\mathrm{ref}}(\vec{p})
    \right|}{f_{\mathrm{ref}}(\vec{p})} \,.
\end{equation}
This value is given for the different angles $\theta$ in \cref{fig:convergence}
(left), as well as a median for each $p$ in \cref{fig:convergence} (right).
Additionally, \cref{fig:convergence} (right) serves as a color guide for most of
the plots presented in this article, as the colors for each $p$ are identical to
the colors in the figures that use color to differentiate between datasets.

Looking at the median representation of the convergence analysis in
\cref{fig:convergence} (right) we see a clear convergence trend for $p \geq 4$,
with decreasing errors as $p$ increases. The discretizations $p = 2$ and $p = 3$
stand out, as the median values deviate from the trend line towards much larger
errors. This can also be seen in \cref{fig:convergence} (left).

We see that for $p = 2$ (the blue line) the angle resolved difference is much
more irregular than, e.g., for $p = 3$ (the orange line). Especially for angles
$\theta \to \SI{90}{\degree}$ we observe many data points that approach very
small values $\Delta f_{p=2}(\vec{p}) \to 0$. This suggests that the signed
numerator in $\Delta f_{p=2}(\vec{p})$, i.e., $f_{p=2}(\vec{p}) -
f_{\mathrm{ref}}(\vec{p})$, undergoes a sign change for these angles. This does
also occur for $p > 2$, however not as frequently. This, together with the
deviation from the trend line indicates that results from a FEM model with $p =
2$ cannot be considered converged.

The angle resolved absolute relative difference for $p = 3$ (the orange line) is
considerably smoother than for $p = 2$. However, similar to $p = 2$, the values
for $\Delta f_{p=3}(\vec{p})$ are at least an order of
magnitude larger than the values for $p \geq 4$. Moreover,
the values $\Delta f_{p=3}(\vec{p})$ are almost always larger than $\Delta
f_{p=2}(\vec{p})$, even though the numerical accuracy parameter is
higher. For a converged model we would expect the opposite.

This behavior forces us to assume that the results obtained from a FEM model
with $p = 2$ and $p = 3$ can not be considered converged. In contrast and
restating what we said above, we consider results obtained with a FEM model
using $p \geq 4$ as converged. Here, increasing the polynomial degree $p$
results in smaller errors, which of course comes at the cost of more numerical
complexity and CPU time spent on a solution.

When we compare the left and right plot of \cref{fig:convergence}, we see a
discrepancy between the \emph{apparent} median of the angle resolved
difference figure and the actually calculated median on the right. This is due
to the non-homogeneous density of the measured angles. The measurement
\emph{Grazing } Incidence X-Ray Fluorescence places a strong emphasis on small
(or grazing) angles. Accordingly, the density of the measured angles is much
greater for angles $\theta \to \SI{90}{\degree}$. This yields a difference
between a median that one would guess at visually, and a median that is actually
calculated.

\subsection{Benchmarks}
\label{subsec:benchmarks}

\begin{figure}[ht]
  \centering
  \includegraphics{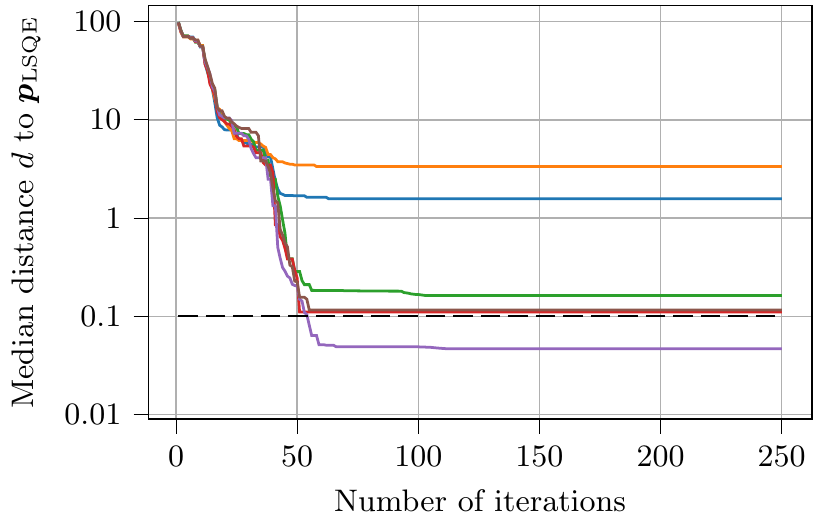}
  \caption{Progress of the parameter reconstruction for different polynomial
    degrees $p$. Shown is the median of six independent optimizations for each
    $p$. The used metric is the distance $d(\vec{p}_{m})$ defined in
    \cref{eq:parameter_diff}. The parameter reconstruction does not succeed when
    the employed model is not converged ($p = 2$ and $p = 3$), as the
    reconstructed parameters do not reach distances of less than one standard
    deviation from the trusted reference value. The parameter reconstruction
    succeeds when using a converged model (here, $p \geq 4$), where generally
    distances of approximately \num{0.1} standard deviations from the trusted
    optimum can be reached. For a color reference we refer the reader to
    \cref{fig:convergence}.}
  \label{fig:optimization_results}
\end{figure}

\begin{figure}[ht]
  \centering
  \includegraphics{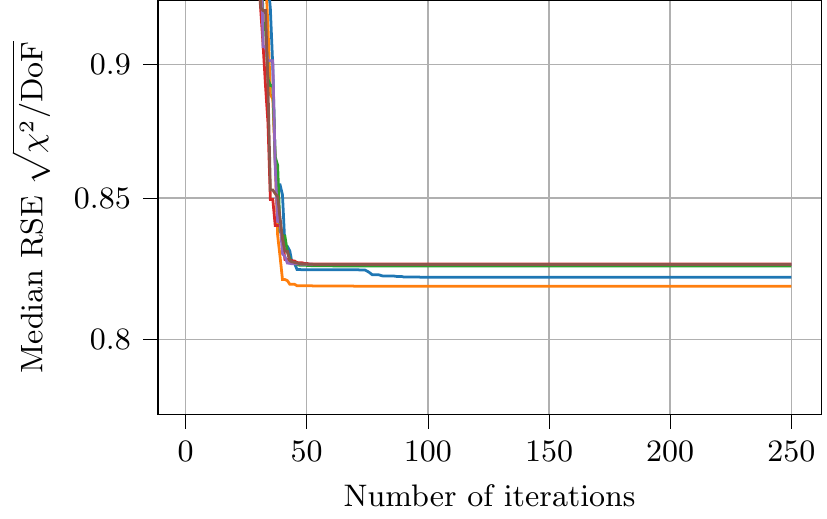}
  \caption{Illustrating example that the regression standard error (RSE) is not
    suited as a metric for comparing parameter reconstructions. The RSE can lead
    to wrong conclusions, as it implicitly assumes that the employed models are
    converged. Conversely, unconverged forward models can show behavior that is
    difficult to predict. Here we see that the model is able to reach a smaller
    RSE with (unconverged) $p = 2$ and $p = 3$. The RSE that can be reached
    using a converged model (i.e., for $p \geq 4$) is slightly larger. For a
    color reference we refer the reader to \cref{fig:convergence}.}
  \label{fig:rse_optimization_results}
\end{figure}

For each of the polynomial degrees $p$ considered in
\cref{subsec:model_convergence} we have performed an optimization benchmark,
where we have run six independent optimizations starting from random starting
points in the parameter space~\cite{plock2022bayesian}. We have provided a
budget of \num{250} iterations to each optimization for reaching the final
reconstructed parameters. This budget was often not exhausted, as many trials
found a good least squares candidate
$\vec{p}_{\mathrm{LSQE}}$ in less than \num{150} iterations.

The metric used to measure convergence is the distance of the reconstructed
parameter at each iteration $m$, $\vec{p}_{m}$, to a -- at the time of the
actual optimization unknown -- reference parameter $\vec{p}_{\mathrm{ref}}$,
where each dimensional component of the compared parameters is weighted by the
-- at optimization time equally unknown -- Gaussian uncertainties
$\vec{\epsilon}_{\mathrm{ref}}$. This distance is calculated as
\begin{equation}
  \label{eq:parameter_diff}
  d(\vec{p}_{m}) = \sqrt{ \sum_{n = 1}^{N}
    \left(
      \frac{\vec{p}_{m,n} - \vec{p}_{\mathrm{ref},n}}{\vec{\epsilon}_{\mathrm{ref},n}}
    \right)^2
  } \,.
\end{equation}
These reference values are chosen to the best reconstructed parameter and the
associated Gaussian uncertainty bands for the most accurate converged FEM model
used in the \emph{optimization benchmarks}, i.e., $\vec{p}_{\mathrm{ref}} =
\vec{p}_{\mathrm{LSQE}}$ and $\vec{\epsilon}_{\mathrm{ref}} =
\vec{\epsilon}_{\mathrm{LSQE}}$ obtained using $p = 7$. ``Best'' in this context
means the parameter with the lowest $\chi^2$ value for that particular $p$.

An obvious alternative metric that is often used when solving
least squares problems is the $\chi^2(\vec{p}_{m})$ value
(or quantities proportional to it such as, e.g., the
regression standard error (RSE)). This metric has the
advantage that it is known immediately at time of optimization. We argue against
using this metric. It can be misleading when performing parameter
reconstructions since it relies on a fully converged model. We show in
\cref{fig:rse_optimization_results} that the unconverged FEM models with $p = 2$
and $p = 3$ actually reach $\chi^2$ values that are marginally smaller than the
values that can be achieved with a fully converged FEM model with $p \geq 4$.
Relying on the $\chi^2$ metric can therefore lead to wrong conclusions and to
falsely attributing importance to unconverged models. This is further visualized
in \cref{subsec:parameter_distributions}.

From the six results for each $p$ we have calculated the distance
$d(\vec{p}_{m})$. For each optimization we have then determined the cumulative
minimum distance and finally calculated the median across each polynomial
degree. These results are shown in \cref{fig:optimization_results}. The color
coding of the figure is identical to the one given in \cref{fig:convergence}
(right).

We observe that the converged models ($p \geq 4$) all reach a distance value $d
\approx \num{0.1}$ standard deviations to the trusted optimum
$\vec{p}_{\mathrm{LSQE}}$. This reproduces the results given in
Reference~\cite{plock2022bayesian}. Here, the median results for
$p = 6$ are clearly below this threshold, while the results for $p \in \left[ 4,
  5, 7 \right]$ remain slightly above it. The median distance results for $p =
2$ and $p = 3$ are only able to reach distance values of $d \gtrapprox 1$. Here,
the results for $p = 3$ are worse than the ones for $p = 2$.

As reference point for all optimizations performed in the benchmark we have
chosen the parameter for $p = 7$ with the lowest $\chi^2$ value. If instead we
choose as reference point for each $p$ the parameter with lowest $\chi^2$ value
for that $p$ (e.g., for $p = 2$ we choose as reference point the best parameter
obtained using $p = 2$), we see some improvements for the median results for $p
= 2$ and $p = 3$. For $p = 3$ the median now also reaches a value of $d \approx
0.1$, and the median for $p = 2$ improves slightly to $d \approx 0.1$. In
\cref{subsec:parameter_uncertainties} we will see that for $p = 3$ this is due
to a shift of the position of the parameter with lowest $\chi^2$ value in the
parameter space. The moderate improvement with $p = 2$ suggests a more complex
function value landscape than for $p \geq 3$.

\subsection{Evaluation of model parameter distributions}
\label{subsec:parameter_distributions}

\begin{figure*}[h]
  \centering
  \includegraphics{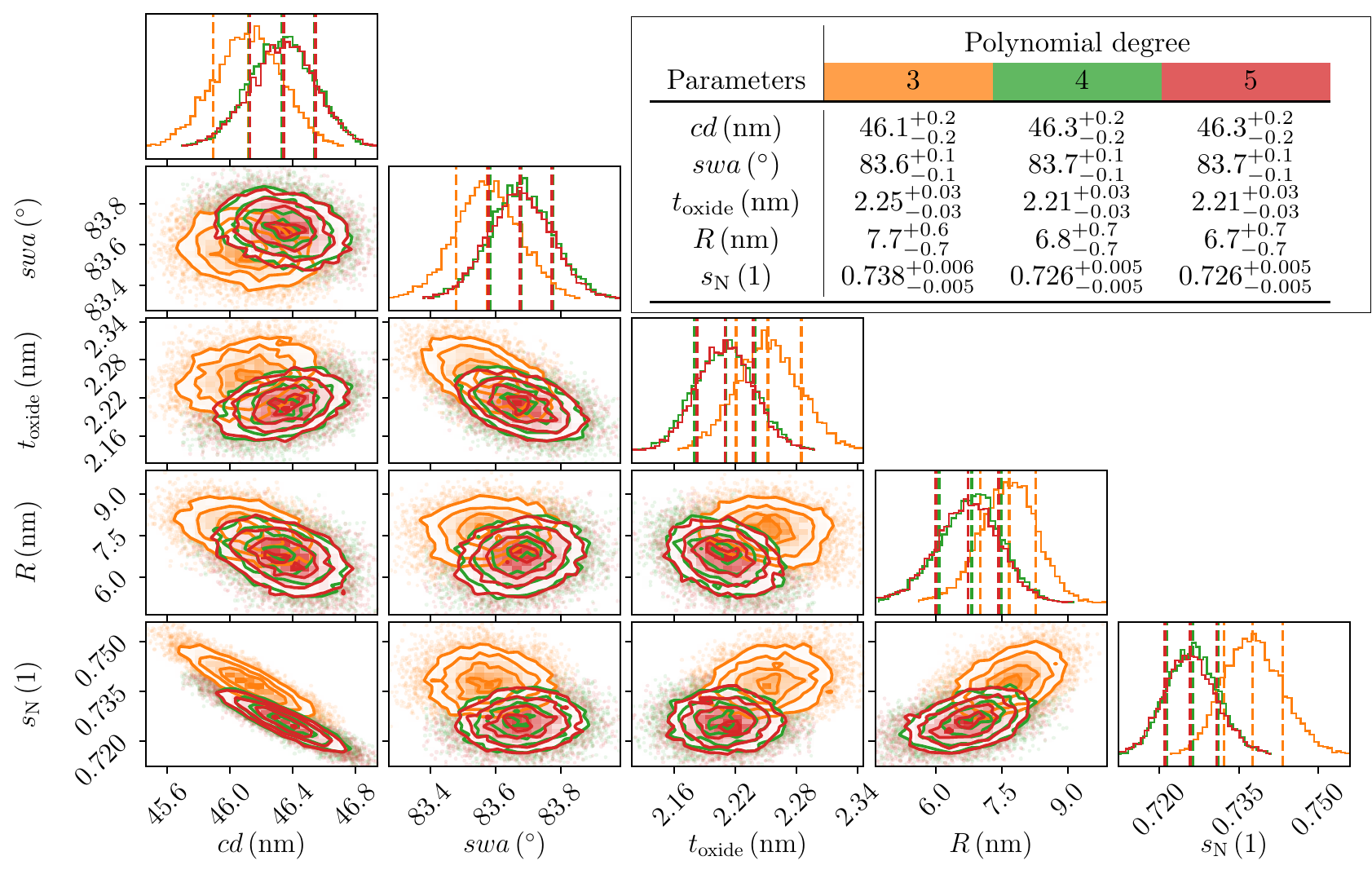}
  \caption{An excerpt of the reconstructed parameters and marginal probability
    distributions for the optimized model function for polynomial degrees $p \in
    \{ 3, 4, 5\}$. For $p = 3$ the model results have not converged yet, this
    leads to a different reconstructed optimum than for a converged model with
    $p \geq 4$. The inset table shows the numerical values for the reconstructed
    parameters and respective marginal distributions. The large numbers are the
    medians of the marginal distributions, while the values in the sub- and
    superscript are the values for the \num{16}'th and \num{84}'th percentile
    respectively.}
  \label{fig:mcmc}
\end{figure*}

Finally, we inspect the function value landscape around the found
least squares estimates for the FEM models with $p \in \{
3, 4, 5\}$ more closely. For this, we use the surrogate aided MCMC approach
detailed in \cref{subsec:parameter_uncertainties} to sample the predicted
likelihood function $\hat{\mathcal{L}}$.

Since the sampled likelihood function in
Reference~\cite{plock2022bayesian} also included a simple error
model, we take the same route to ensure comparable results. Therefore, we also
determined the distribution of the parameter $\eta_{\mathrm{c}}$ in the error
model
\begin{equation}
  \label{eq:simple_error_model}
  \tilde{\eta}_{k}(\eta_{\mathrm{c}}) = 2^{\eta_{\mathrm{c}}}\cdot\eta_{k}
\end{equation} during the sampling process. This error model allows to scale the
measurement errors of the experiment up or down. Accordingly, we draw samples
from the predicted likelihood function
\begin{equation}
  \label{eq:predicted_least_square_likelihood}
  \overline{\mathcal{L}}(\vec{p}, \eta_{\mathrm{c}}) = \prod_{k = 1}^{K} \frac{1}{\sqrt{2\pi\tilde{\eta}_{k}^2(\eta_{\mathrm{c}})}}
  \exp{ \left( -
      \frac{1}{2} \frac{\left( \overline{y}_{k}(\vec{p}) - t_{k} \right) ^2}{\tilde{\eta}_{k}^{2}(\eta_{\mathrm{c}})} \right) } \,.
\end{equation}
We omit $p = 2$ from this analysis because we showed that already the model
for $p = 3$ is not converged, as well as $p \geq 6$, because we will already see
very little difference between the model values for $p = 4$ and $p = 5$.

After the optimization benchmark using the BTVO we have six pre-trained
surrogate models for each of the polynomial orders $p$ considered. These
surrogate models were trained on data originating from simulations of the
respective FEM model; for a detailed discussion on the model and the parameters
used we refer the reader to Reference~\cite{plock2022bayesian}.
The surrogates can be used as cheap-to-evaluate accurate approximations of the
expensive FEM model. Due to the random optimization starting points each of the
surrogates contains different samples from the parameter space, that were
investigated during the optimizations. For each $p \in \{3, 4, 5\}$ we select
the surrogate model that has led to the smallest $\chi^2$ value, and use this
surrogate model as a starting point for the refinement process. For this
pre-sampling refinement stage we provided a refinement budget of \num{250}
additional samples that were used to increase the fidelity of the surrogate
models in the proximity of the different $\vec{p}_{\mathrm{LSQE}}$. After
refinement, we then employed \texttt{emcee}~\cite{emcee} to construct \num{32}
Markov chains containing \num{50000} samples each, for the three $p$
investigated.

The MCMC sampling region was restricted to a boxed region of three standard
deviations $\vec{\epsilon}_{\mathrm{LSQE}}$ around the
least squares estimate $\vec{p}_{\mathrm{LSQE}}$. This was
done because even after the refinement stage we can only assume that the
surrogate model is sufficiently refined in a region close to the
least squares estimate. In regions far away from training
data the surrogate predictions revert back to the priors used in the
construction of the GPs. This leads to difficult-to-predict behavior of the
sampled predicted likelihood function, such that the MCMC sampler may falsely
attribute importance to undersampled regions and therefore giving misleading
results.

The results for a small subset of the marginal distributions of the
reconstructed parameter are given in \cref{fig:mcmc}. Therein we also give
tabulated the median, as well as lower and upper standard deviation of the marginal
model parameter distribution for the shown parameters and polynomial degrees
$p$. The \emph{complete} list of medians and lower/upper standard deviations, as
well as the results for the fit error model, is given in \cref{tab:percentiles}.
Note that for most of the additional parameters given in \cref{tab:percentiles}
we report similar results (within the reconstructed uncertainties) for all $p$
considered.

For the displayed parameters we see that the model parameter distributions for
the converged models ($p = 4$, and $p = 5$) are identical within the
reconstructed model parameter uncertainties. In contrast to that, using an
unconverged FEM model with $p = 3$ leads to a shift of more than one standard
deviation of the median of the model parameter distribution for several
components of the model parameter $\vec{p}$. The maximum likelihood estimate
obtained by MCMC sampling for the parameter $cd$ and $swa$ with $p = 3$ are
approximately one standard deviation smaller than the ones obtained with $p \geq
4$. Conversely, the model parameters $t_{\mathrm{oxide}}$, $R$, and
$s_{\mathrm{N}}$ are approximately one standard deviation larger. This explains
why the median distance of the results for $p = 3$ in
\cref{fig:optimization_results} remain a distance of more than one standard
deviation away from the $d \approx 0.1$ threshold. Similarly, it also explains
why the use of an alternative reference value leads to $d \approx 0.1$ for $p =
3$, since only the position of the best reconstruction value appears to have
shifted, while the general shape of the model parameter distribution is
unaltered. The reconstructed uncertainties of the model parameters appear mostly
unaffected by the use of an unconverged FEM model. Here, only the parameters $R$
and $t_{\mathrm{Sub}}$ appear to have gained a small skewedness when using $p =
3$.

When using a converged FEM model we report the same model parameter distribution
as Reference~\cite{plock2022bayesian} for the small subset given
therein.

Finally, w.r.t. the fit parameter $\eta_{\mathrm{c}}$ of the simple error model
\cref{eq:simple_error_model}, we did not observe an impact of the proper or
improper choice of the polynomial degree $p$ on an over- or underestimation of
the measurement errors.

\subsection{Discussion}
\label{subsec:discussion}

For the problem considered in this article the modeling differences between the
polynomial degrees $p = 3$ and $p \geq 4$, described in
\cref{subsec:benchmarks,subsec:model_convergence,subsec:parameter_distributions},
are mostly seen as a shift in the distribution. Here, the median is shifted to a
different position and a small skewedness is introduced to some of the model
parameter marginal distributions.

We hypothesize that the shift in the median value for the different model
accuracies is due to a very coarse sampling of the fields of Maxwells equations
in the oxide layer of the mesh at the reconstructed model parameters. In
simulations with a minimum mesh element size of $h = \SI{5}{\nano\meter}$ and an
oxide layer thickness of $t_{\mathrm{oxide}} \approx \SI{2.2}{\nano\meter}$,
this oxide layer will only be one element thick. Therefore, an already very
small number of unknowns are reserved for calculating the effect of it on the
simulated fluorescence signal. At the same time this oxide
layer is arguably among the most important components of the model, since it is
the first part of the structure that the electromanegtic fields interact with.

This is exacerbated by the fact that any (even just minute) rotation of the
sample during the measurement process will result in it being illuminated only
from the top and the side. Additionally, for small (\emph{grazing}) incidence
angles one can assume that only the upper side of the structure is illuminated,
since much of the incidence radiation is captured by other parts of the grating.
In general, the bottom of the grating will not be illuminated strongly.

All of this is not captured in the model, but is contained in the experimental
measurement. As such, the model parameters that control the top and top side of
the structure will have a great importance on the reconstruction process. This
is further exacerbated by the strong emphasis on grazing incidence angles
$\theta \to \SI{90}{\degree}$ during the experiment itself (by performing more
experimental measurements there), and therefore for the
least squares estimate.

The parameters that control the top and top side portion of the gratings unit
cell structure are therefore very influential for the parameter reconstruction
process. Here we have seen the largest difference in the reconstructed
parameters. The two different parameterizations seen in \cref{fig:mcmc} and
\cref{tab:percentiles}, i.e., (when going from $p = 3$ to $p = 4$) reducing the
critical dimension $cd$ and side wall angle $swa$ while increasing the radius
$R$ and the thickness of the oxide layer $t_{\mathrm{oxide}}$, approximately
keep the amount of material in the oxide layer constant. In their respective
accuracy setting of $p = 3$ and $p \geq 4$ they both apparently yield similar
physical behavior, despite being several standard deviations (in reconstructed
uncertainties) apart from each other.

\section{Conclusion}
\label{sec:conclusion}

We have presented an analysis of the impact of a properly or improperly chosen
polynomial degrees $p$ in a finite element method (FEM) model
during a parameter reconstruction. This polynomial degree determines the
accuracy, and as such the convergence, of the FEM model.

We have investigated the $p$ convergence of the FEM model and identified
different $p$ for which the model results can be assumed to be converged, but
also $p$ for which we can expect model behavior that is difficult to predict.

This difficult to predict behavior had manifested itself in apparently better
reconstruction results, especially when using inappropriate parameter
reconstruction metrics such as the regression standard error.

Finally, we were able to make use of a surrogate aided MCMC approach to show
that in this particular FEM model the improperly chosen polynomial degree $p =
3$ resulted in a shift of the parameter reconstruction result from the converged
parameter reconstruction result for $p \geq 4$.

This article highlights the importance of using properly converged models in
parameter reconstructions. It also stresses that ignorance about the state of
model convergence can result in falsely attributing importance to unconverged
results.

\section{Conflicts of Interest}

The authors declare no conflict of interest.

\section{Acknowledgements}

This project has received funding from the German Federal Ministry of Education
and Research (BMBF, project number 05M20ZAA, siMLopt; project number 01IS20080A,
SiM4diM; Forschungscampus MODAL, project number 05M20ZBM). This project has
received funding from the EMPIR programme co-financed by the Participating
States and by the European Union’s Horizon 2020 research and innovation
programme (project 20IND04 ”ATMOC”; project 20FUN02 ”POLIGHT”). The authors
gratefully acknowledge the scientific support and HPC resources provided by the
NHR center at ZIB. NHR funding is provided by federal and Berlin state
authorities.

\appendix

\section{Table with complete reconstructed distribution}
\label{sec:complete_dist_table}

\cref{tab:percentiles} shows a listing of the percentiles of the full model
parameter distribution for the polynomial degrees $p \in \{ 3, 4, 5 \}$.

\begin{table*}[h]
  \centering
  \begin{tabular}{c|ccc}
    &\multicolumn{3}{c}{Polynomial degree}\\
    Parameters & \cellcolor{darkorange25512714!75} 3 & \cellcolor{forestgreen4416044!75} 4 & \cellcolor{crimson2143940!75} 5 \\
    \toprule

    $h\,(\mathrm{nm})$ & ${\num{89.6}}_{-\num{0.4}}^{+\num{0.4}}$ & ${\num{89.5}}_{-\num{0.4}}^{+\num{0.4}}$ & ${\num{89.5}}_{-\num{0.4}}^{+\num{0.4}}$ \\
    $cd\,(\mathrm{nm})$ & ${\num{46.1}}_{-\num{0.2}}^{+\num{0.2}}$ & ${\num{46.3}}_{-\num{0.2}}^{+\num{0.2}}$ & ${\num{46.3}}_{-\num{0.2}}^{+\num{0.2}}$ \\
    $swa\,(^\circ)$ & ${\num{83.6}}_{-\num{0.1}}^{+\num{0.1}}$ & ${\num{83.7}}_{-\num{0.1}}^{+\num{0.1}}$ & ${\num{83.7}}_{-\num{0.1}}^{+\num{0.1}}$ \\
    $t_{\mathrm{oxide}}\,(\mathrm{nm})$ & ${\num{2.25}}_{-\num{0.03}}^{+\num{0.03}}$ & ${\num{2.21}}_{-\num{0.03}}^{+\num{0.03}}$ & ${\num{2.21}}_{-\num{0.03}}^{+\num{0.03}}$ \\
    $t_{\mathrm{groove}}\,(\mathrm{nm})$ & ${\num{1.1}}_{-\num{0.3}}^{+\num{0.3}}$ & ${\num{1.1}}_{-\num{0.3}}^{+\num{0.3}}$ & ${\num{1.0}}_{-\num{0.3}}^{+\num{0.3}}$ \\
    $R\,(\mathrm{nm})$ & ${\num{7.7}}_{-\num{0.7}}^{+\num{0.6}}$ & ${\num{6.8}}_{-\num{0.7}}^{+\num{0.7}}$ & ${\num{6.7}}_{-\num{0.7}}^{+\num{0.7}}$ \\
    $t_{\mathrm{Sub}}\,(\mathrm{nm})$ & ${\num{6.8}}_{-\num{0.9}}^{+\num{0.8}}$ & ${\num{7.0}}_{-\num{0.9}}^{+\num{0.8}}$ & ${\num{7.0}}_{-\num{0.9}}^{+\num{0.8}}$ \\
    $s_{\mathrm{N}}\,(1)$ & ${\num{0.738}}_{-\num{0.005}}^{+\num{0.006}}$ & ${\num{0.726}}_{-\num{0.005}}^{+\num{0.005}}$ & ${\num{0.726}}_{-\num{0.005}}^{+\num{0.005}}$ \\
    $\Delta_{{\theta}}\,(^\circ)$ & ${\num{-0.099}}_{-\num{0.003}}^{+\num{0.003}}$ & ${\num{-0.102}}_{-\num{0.003}}^{+\num{0.003}}$ & ${\num{-0.102}}_{-\num{0.003}}^{+\num{0.003}}$ \\
    $\Delta_{{\phi}}\,(^\circ)$ & ${\num{0.003}}_{-\num{0.009}}^{+\num{0.01}}$ & ${\num{0.007}}_{-\num{0.009}}^{+\num{0.011}}$ & ${\num{0.007}}_{-\num{0.01}}^{+\num{0.011}}$ \\
    $\eta_{c}\,(1)$ & ${\num{0.002}}_{-\num{0.07}}^{+\num{0.074}}$ & ${\num{0.002}}_{-\num{0.07}}^{+\num{0.073}}$ & ${\num{0.002}}_{-\num{0.072}}^{+\num{0.073}}$ \\

    \bottomrule
  \end{tabular}
  \caption{A complete listing of the reconstructed model parameters and their
    distribution percentiles as determined by Markov chain Monte Carlo sampling
    of the refined surrogate model of the expensive model function shown in
    \cref{fig:mcmc}. The color coding of the polynomial degree title cells
    matches the colors of the figures in the main article.}
  \label{tab:percentiles}
\end{table*}

\section{Table with run times of the forward problem}
\label{sec:runtimes}

In each evaluation of the forward problem detailed in \cref{subsec:model}, 208
discrete calculations have to be performed. The run times of these solutions
crucially depend on the chosen accuracy level, which is here controlled by means
of the polynomial degree $p$ of the finite elements in the problem
discretization. In \cref{tab:runtimes} we give a listing of these run times,
where we differentiate between the parallelized run time (or wall time), the
single thread run time (where we are summing up all 208 individual run times),
and the median run time (together with standard deviations obtained by means of
the 16'th and 84'th percentile of the 208 individual run times). Please note
that due to the perfectly parallel nature of the forward problem (the 208
discrete incidence angles can be computed independent of each other) the time
given in the ``Parallelized run time'' column will vary depending on the
available resources. The times given in \cref{tab:runtimes} were determined on a
workstation computer with an AMD Ryzen 7 3700X 8-Core processor (yielding 16
workable threads) with 32 GB of RAM installed. Here, 12 threads were reserved
for solving the forward problem such that the remaining 4 threads could be
utilized to deal with the remaining tasks running on the computer (e.g.
operating system, internet browser, etc).

Consider that when no surrogate is used to lessen the computational impact, each
data point in each chain of the Markov chain Monte Carlo sampling strategy
requires approximately the time given in the ``Parallelized run time'' column to
compute.

\begin{table*}[h]
  \centering
  \begin{tabular}{cccc}
    FEM degree $p$ & Parallelized run time & Single thread run time & Median run time \\
    \toprule
    $8$ ($p_{\mathrm{ref}}$) & $\SI{256.7}{\second}$ & $\SI{2895.3}{\second}$ & $\SI{13.8}{\second}_{-\SI{2.0}{\second}}^{+\SI{1.7}{\second}}$ \\
    $7$ & $\SI{170.1}{\second}$ & $\SI{1871.7}{\second}$ & $\SI{8.8}{\second}_{-\SI{1.1}{\second}}^{+\SI{1.3}{\second}}$ \\
    $6$ & $\SI{126.3}{\second}$ & $\SI{1359.3}{\second}$ & $\SI{6.4}{\second}_{-\SI{0.6}{\second}}^{+\SI{0.8}{\second}}$ \\
    $5$ & $\SI{86.6}{\second}$ & $\SI{891.4}{\second}$ & $\SI{4.2}{\second}_{-\SI{0.5}{\second}}^{+\SI{0.5}{\second}}$ \\
    $4$ & $\SI{66.7}{\second}$ & $\SI{651.2}{\second}$ & $\SI{3.1}{\second}_{-\SI{0.3}{\second}}^{+\SI{0.3}{\second}}$ \\
    $3$ & $\SI{47.6}{\second}$ & $\SI{445.5}{\second}$ & $\SI{2.2}{\second}_{-\SI{0.2}{\second}}^{+\SI{0.1}{\second}}$ \\
    $2$ & $\SI{36.1}{\second}$ & $\SI{312.7}{\second}$ & $\SI{1.5}{\second}_{-\SI{0.1}{\second}}^{+\SI{0.1}{\second}}$ \\
    \bottomrule
  \end{tabular}
  \caption{ A listing of the run times for the different accuracy levels of the
    forward problem. In each evaluation of the forward problem, 208 single
    calculations have to be performed. The parallelized run times (or wall
    times) for a single evaluation was obtained under the conditions given in
    \cref{sec:runtimes}. Single thread run times are the sum of all 208
    individual run times. Median run times are the 50'th percentile of all 208
    run times. Lower and upper standard deviations are determined from the 16'th
    and 84'th percentiles of all 208 run times.}
  \label{tab:runtimes}
\end{table*}

\bibliographystyle{IEEEtran}
\bibliography{bibliography}

\begin{thebibliography}{10}
\providecommand{\url}[1]{#1}
\csname url@samestyle\endcsname
\providecommand{\newblock}{\relax}
\providecommand{\bibinfo}[2]{#2}
\providecommand{\BIBentrySTDinterwordspacing}{\spaceskip=0pt\relax}
\providecommand{\BIBentryALTinterwordstretchfactor}{4}
\providecommand{\BIBentryALTinterwordspacing}{\spaceskip=\fontdimen2\font plus
\BIBentryALTinterwordstretchfactor\fontdimen3\font minus
  \fontdimen4\font\relax}
\providecommand{\BIBforeignlanguage}[2]{{%
\expandafter\ifx\csname l@#1\endcsname\relax
\typeout{** WARNING: IEEEtran.bst: No hyphenation pattern has been}%
\typeout{** loaded for the language `#1'. Using the pattern for}%
\typeout{** the default language instead.}%
\else
\language=\csname l@#1\endcsname
\fi
#2}}
\providecommand{\BIBdecl}{\relax}
\BIBdecl

\bibitem{diebold2013nanoscale}
A.~C. Diebold, ``Nanoscale characterization and metrology,'' \emph{J. Vac. Sci.
  Technol. A}, vol.~31, p. 050804, 2013.

\bibitem{orji2018metrology}
N.~G. Orji, M.~Badaroglu, B.~M. Barnes, C.~Beitia, B.~D. Bunday, U.~Celano,
  R.~J. Kline, M.~Neisser, Y.~Obeng, and A.~Vladar, ``Metrology for the next
  generation of semiconductor devices,'' \emph{Nat. Electron.}, vol.~1, pp.
  532--547, 2018.

\bibitem{denboef2016optical}
A.~J. den Boef, ``Optical wafer metrology sensors for process-robust {CD} and
  overlay control in semiconductor device manufacturing,'' \emph{Surf.
  Topogr.}, vol.~4, p. 023001, 2016.

\bibitem{endres2014investigations}
J.~Endres, A.~Diener, M.~Wurm, and B.~Bodermann, ``Investigations of the
  influence of common approximations in scatterometry for dimensional
  nanometrology,'' \emph{Meas. Sci. Technol.}, vol.~25, p. 044004, 2014.

\bibitem{jones2003small}
R.~L. Jones, T.~Hu, E.~K. Lin, W.-L. Wu, R.~Kolb, D.~M. Casa, P.~J. Bolton, and
  G.~G. Barclay, ``Small angle {X}-ray scattering for sub-100 nm pattern
  characterization,'' \emph{Appl. Phys. Lett.}, vol.~83, p. 4059, 2003.

\bibitem{omullane2016modeling}
S.~O’Mullane, N.~Keller, and A.~C. Diebold, ``Modeling ellipsometric
  measurement of three-dimensional structures with rigorous coupled wave
  analysis and finite element method simulations,'' \emph{J. Micro.
  Nanolithogr. MEMS MOEMS}, vol.~15, p. 044003, 2016.

\bibitem{attota2016feasibility}
R.~K. Attota, P.~Weck, J.~A. Kramar, B.~Bunday, and V.~Vartanian, ``Feasibility
  study on {3-D} shape analysis of high-aspect-ratio features using
  through-focus scanning optical microscopy,'' \emph{Opt. Express}, vol.~24, p.
  16574, 2016.

\bibitem{Soltwisch2016prb}
V.~Soltwisch, A.~Haase, J.~Wernecke, J.~Probst, M.~Schoengen, S.~Burger,
  M.~Krumrey, and F.~Scholze, ``Correlated diffuse x-ray scattering from
  periodically nanostructured surfaces,'' \emph{Phys. Rev. B}, vol.~94, p.
  035419, 2016.

\bibitem{aster2018parameter}
R.~C. Aster, B.~Borchers, and C.~H. Thurber, \emph{Parameter estimation and
  inverse problems}.\hskip 1em plus 0.5em minus 0.4em\relax Elsevier, 2018.

\bibitem{Hamm:17}
M.~Hammerschmidt, M.~Weiser, X.~G. Santiago, L.~Zschiedrich, B.~Bodermann, and
  S.~Burger, ``{Quantifying parameter uncertainties in optical scatterometry
  using Bayesian inversion},'' \emph{Proc. SPIE}, vol. 10330, p. 1033004, 2017.

\bibitem{STORCH20071417}
R.~B. Storch, L.~C. Pimentel, and H.~R. Orlande, ``Identification of
  atmospheric boundary layer parameters by inverse problem,'' \emph{Atmospheric
  Environment}, vol.~41, no.~7, pp. 1417--1425, 2007.

\bibitem{Schn:2019Benchmark}
P.-I. Schneider, X.~Garcia~Santiago, V.~Soltwisch, M.~Hammerschmidt, S.~Burger,
  and C.~Rockstuhl, ``Benchmarking five global optimization approaches for
  nano-optical shape optimization and parameter reconstruction,'' \emph{ACS
  Photonics}, vol.~6, no.~11, pp. 2726--2733, 2019.

\bibitem{PSO_magnetotelluric}
F.~Pace, A.~Santilano, and A.~Godio, ``{Particle swarm optimization of 2D
  magnetotelluric data},'' \emph{Geophysics}, vol.~84, no.~3, pp. E125--E141,
  03 2019.

\bibitem{schwaab2008nonlinear}
M.~Schwaab, E.~C. Biscaia~Jr, J.~L. Monteiro, and J.~C. Pinto, ``Nonlinear
  parameter estimation through particle swarm optimization,'' \emph{Chemical
  Engineering Science}, vol.~63, no.~6, pp. 1542--1552, 2008.

\bibitem{lobato2012estimation}
F.~Lobato, V.~Steffen~Jr, and A.~S. Neto, ``Estimation of space-dependent
  single scattering albedo in a radiative transfer problem using differential
  evolution,'' \emph{Inverse Problems in Science and Engineering}, vol.~20,
  no.~7, pp. 1043--1055, 2012.

\bibitem{cavalini2016model}
A.~A. Cavalini~Jr, F.~S. Lobato, E.~H. Koroishi, and V.~Steffen~Jr, ``Model
  updating of a rotating machine using the self-adaptive differential evolution
  algorithm,'' \emph{Inverse Problems in Science and Engineering}, vol.~24,
  no.~3, pp. 504--523, 2016.

\bibitem{Herrero:21}
A.~F. Herrero, M.~Pfl\"{u}ger, J.~Puls, F.~Scholze, and V.~Soltwisch,
  ``Uncertainties in the reconstruction of nanostructures in {EUV}
  scatterometry and grazing incidence small-angle {X}-ray scattering,''
  \emph{Opt. Express}, vol.~29, pp. 35\,580--35\,591, 2021.

\bibitem{williams2006gaussian}
C.~K. Williams and C.~E. Rasmussen, \emph{Gaussian processes for machine
  learning}.\hskip 1em plus 0.5em minus 0.4em\relax MIT Press, 2006.

\bibitem{plock2022bayesian}
M.~Plock, K.~Andrle, S.~Burger, and P.-I. Schneider, ``Bayesian target-vector
  optimization for efficient parameter reconstruction,'' \emph{Adv. Theory
  Simul.}, vol.~5, no.~10, p. 2200112, 2022.

\bibitem{andrle2021shape}
A.~Andrle, P.~Hönicke, G.~Gwalt, P.-I. Schneider, Y.~Kayser, F.~Siewert, and
  V.~Soltwisch, ``{Shape- and Element-Sensitive Reconstruction of Periodic
  Nanostructures with Grazing Incidence X-ray Fluorescence Analysis and Machine
  Learning},'' \emph{Nanomaterials}, vol.~11, p. 1647, 2021.

\bibitem{martino2021automatic}
L.~Martino, F.~Llorente, E.~Curbelo, J.~L{\'o}pez-Santiago, and J.~M{\'\i}guez,
  ``Automatic tempered posterior distributions for bayesian inversion
  problems,'' \emph{Mathematics}, vol.~9, no.~7, p. 784, 2021.

\bibitem{leve:1944}
K.~Levenberg, ``A method for the solution of certain non-linear problems in
  least squares,'' \emph{Q. Appl. Math.}, vol.~2, no.~2, pp. 164--168, 1944.

\bibitem{marq:1963}
D.~W. Marquardt, ``An algorithm for least-squares estimation of nonlinear
  parameters,'' \emph{J. Soc. Indust. Appl. Math.}, vol.~11, no.~2, pp.
  431--441, 1963.

\bibitem{flet:1971}
R.~Fletcher, ``A modified {M}arquardt subroutine for non-linear least
  squares,'' Atomic Energy Research Establishment, Harwell (England), Tech.
  Rep. {AERE-R-6799}, 1971.

\bibitem{kutner2005applied}
M.~H. Kutner, C.~J. Nachtsheim, J.~Neter, and W.~Li, \emph{Applied Linear
  Statistical Models}, 5th~ed.\hskip 1em plus 0.5em minus 0.4em\relax
  MCGraw-Hill Irwin, 2005.

\bibitem{strutz2010data}
T.~Strutz, \emph{Data Fitting and Uncertainty: A Practical Introduction to
  Weighted Least Squares and Beyond}.\hskip 1em plus 0.5em minus 0.4em\relax
  Vieweg and Teubner, 2011.

\bibitem{press2007numerical}
W.~H. Press, S.~A. Teukolsky, W.~T. Vetterling, and B.~P. Flannery,
  \emph{Numerical Recipes: The Art of Scientific Computing}, 3rd~ed.\hskip 1em
  plus 0.5em minus 0.4em\relax Cambridge University Press, 2007.

\bibitem{brochu2010tutorial}
E.~Brochu, V.~M. Cora, and N.~de~Freitas, ``A tutorial on {B}ayesian
  optimization of expensive cost functions, with application to active user
  modeling and hierarchical reinforcement learning,'' \emph{arXiv preprint,
  arXiv:1012.2599}, 2010.

\bibitem{jones1998efficient}
D.~R. Jones, M.~Schonlau, and W.~J. Welch, ``Efficient global optimization of
  expensive black-box functions,'' \emph{J. Global Optim.}, vol.~13, pp.
  455--492, 1998.

\bibitem{Schn:2019GPR}
P.-I. Schneider, M.~Hammerschmidt, L.~Zschiedrich, and S.~Burger, ``Using
  {Gaussian} process regression for efficient parameter reconstruction,''
  \emph{Proc. SPIE}, vol. 10959, p. 1095911, 2019.

\bibitem{garnett_bayesoptbook_2023}
R.~Garnett, \emph{Bayesian Optimization}.\hskip 1em plus 0.5em minus
  0.4em\relax Cambridge University Press, 2023.

\bibitem{byrd1995limited}
R.~H. Byrd, P.~Lu, J.~Nocedal, and C.~Zhu, ``A limited memory algorithm for
  bound constrained optimization,'' \emph{SIAM J. Sci. Comp.}, vol.~16, no.~5,
  pp. 1190--1208, 1995.

\bibitem{deuflhard2005newton}
P.~Deuflhard, \emph{Newton methods for nonlinear problems: affine invariance
  and adaptive algorithms}.\hskip 1em plus 0.5em minus 0.4em\relax Springer
  Science \& Business Media, 2005, vol.~35.

\bibitem{uhrenholt}
A.~K. Uhrenholt and B.~S. Jensen, ``Efficient {Bayesian} optimization for
  target vector estimation,'' in \emph{The 22nd International Conference on
  Artificial Intelligence and Statistics}.\hskip 1em plus 0.5em minus
  0.4em\relax PMLR, 2019, pp. 2661--2670.

\bibitem{alvarez2011kernels}
M.~A. Alvarez, L.~Rosasco, and N.~D. Lawrence, ``Kernels for vector-valued
  functions: {A} review,'' \emph{arXiv preprint arXiv:1106.6251}, 2011.

\bibitem{liu2018remarks}
H.~Liu, J.~Cai, and Y.-S. Ong, ``Remarks on multi-output {G}aussian process
  regression,'' \emph{Knowl.-Based Syst.}, vol. 144, pp. 102--121, 2018.

\bibitem{matsui2019bayesian}
K.~Matsui, S.~Kusakawa, K.~Ando, K.~Kutsukake, T.~Ujihara, and I.~Takeuchi,
  ``Bayesian active learning for structured output design,'' \emph{arXiv
  preprint, arXiv:1911.03671}, 2019.

\bibitem{mathai1992quadratic}
A.~M. Mathai and S.~B. Provost, \emph{Quadratic forms in random variables:
  theory and applications}.\hskip 1em plus 0.5em minus 0.4em\relax Dekker,
  1992.

\bibitem{mohsenipour2012distribution}
A.~A. Mohsenipour, ``On the distribution of quadratic expressions in various
  types of random vectors,'' Ph.D. dissertation, University of Western Ontario,
  2012.

\bibitem{schneider2022reconstructing}
P.-I. Schneider, P.~Manley, J.~Kr{\"u}ger, L.~Zschiedrich, R.~K{\"o}ning,
  B.~Bodermann, and S.~Burger, ``Reconstructing phase aberrations for
  high-precision dimensional microscopy,'' \emph{Proc. SPIE}, vol. 12137, p.
  121370I, 2022.

\bibitem{garcia2018shape}
X.~Garcia-Santiago, P.-I. Schneider, C.~Rockstuhl, and S.~Burger, ``Shape
  design of a reflecting surface using bayesian optimization,'' \emph{J. Phys.
  Conf. Ser.}, vol. 963, p. 012003, 2018.

\bibitem{andrieu2003introduction}
C.~Andrieu, N.~De~Freitas, A.~Doucet, and M.~I. Jordan, ``An introduction to
  {MCMC} for machine learning,'' \emph{Mach. Learn.}, vol.~50, no.~1, pp.
  5--43, 2003.

\bibitem{sammut2011encyclopedia}
C.~Sammut and G.~I. Webb, \emph{Encyclopedia of machine learning}.\hskip 1em
  plus 0.5em minus 0.4em\relax Springer Science \& Business Media, 2011.

\bibitem{friedman2001elements}
J.~Friedman, T.~Hastie, R.~Tibshirani \emph{et~al.}, \emph{The elements of
  statistical learning}.\hskip 1em plus 0.5em minus 0.4em\relax Springer series
  in statistics New York, 2001, vol.~1, no.~10.

\bibitem{degroot2012probability}
M.~H. DeGroot and M.~J. Schervish, \emph{Probability and statistics},
  4th~ed.\hskip 1em plus 0.5em minus 0.4em\relax Pearson, 2012.

\bibitem{emcee}
D.~{Foreman-Mackey}, D.~W. {Hogg}, D.~{Lang}, and J.~{Goodman}, ``{emcee: The
  MCMC Hammer},'' \emph{Publ. Astron. Soc. Pac.}, vol. 125, no. 925, p. 306,
  2013.

\bibitem{cornerpy}
D.~Foreman-Mackey, ``corner.py: Scatterplot matrices in python,'' \emph{The
  Journal of Open Source Software}, vol.~1, no.~2, p.~24, 2016.

\bibitem{Monk2003a}
P.~Monk, \emph{Finite Element Methods for {M}axwell's Equations}.\hskip 1em
  plus 0.5em minus 0.4em\relax Oxford: Claredon Press, 2003.

\bibitem{jcm_scattering}
J.~Pomplun, S.~Burger, L.~Zschiedrich, and F.~Schmidt, ``Adaptive finite
  element method for simulation of optical nano structures,'' \emph{Phys.
  Status Solidi B}, vol. 244, no.~10, p. 3419, 2007.

\bibitem{Burger2008ipnra}
S.~Burger, L.~Zschiedrich, J.~Pomplun, and F.~Schmidt, ``{JCMsuite}: {A}n
  adaptive {FEM} solver for precise simulations in nano-optics,'' in
  \emph{Integrated Photonics and Nanophotonics Research and
  Applications}.\hskip 1em plus 0.5em minus 0.4em\relax Optical Society of
  America, 2008, p. ITuE4.

\bibitem{C8NR00328A}
V.~Soltwisch, P.~Hönicke, Y.~Kayser, J.~Eilbracht, J.~Probst, F.~Scholze, and
  B.~Beckhoff, ``Element sensitive reconstruction of nanostructured surfaces
  with finite elements and grazing incidence soft {X}-ray fluorescence,''
  \emph{Nanoscale}, vol.~10, pp. 6177--6185, 2018.

\end{thebibliography}

\end{document}